\documentclass{article}

\usepackage{graphicx}

\begin{document}

\title{Bessel beam through a dielectric slab at oblique incidence:\\ the case of
 total reflection}
\author{D. Mugnai\footnote{d.mugnai@ifac.cnr.it} \\
{\it ``Nello Carrara'' Institute of Applied Physics - CNR
\footnote{Former: ``Nello Carrara" Electromagnetic Waves Research
Institute} }
\\ {\it 50127 Firenze, Italy}}

\maketitle

\date{}

\begin{abstract}
The oblique incidence of a Bessel beam on a dielectric slab
with  refractive index $n_1$
surrounded by a medium of a refractive index $n>n_1$ may be
studied simply by expanding the Bessel beam into a set of plane waves forming
the same angle
$\theta_0$ with the axis of the beam.
In the present paper we examine
a Bessel beam that impinges at oblique incidence
onto a layer in such a way that each plane-wave
component impinges with an angle larger than the critical angle.
\end{abstract}

 \vspace{0.5 cm}

The propagation of a Bessel beam, or Bessel wave\cite{note0},
represents an interesting topic both for its implication in relation to
superluminal motion\cite{saa,mug-prl}, and for its spatial
localization, which makes it a good candidate
for all practical applications where a localized field is
required\cite{tor0,tay}.

The propagation of a Bessel beam through a dielectric slab,
for normal incidence and in the presence of total reflection,
has already been analyzed by Shaarawi and Besieris\cite{sha} and Mugnai\cite{mug01}.
Furthermore , propagation in a layered medium for normal and oblique
incidences, in the absence of total reflection, has also been
studied\cite{note}.

The aim of the present work is to analyze a Bessel beam
impinging onto a plane slab at oblique incidence,
in the case in which all the waves forming the beam are in total reflection.
The main interest of this kind of analysis is to investigate if the beam propagates by maintaining its characteristic of localized wave also in the case of oblique incidence.

Let us consider a simple system formed by two half-spaces
that are filled with a homogeneous and non-dispersive medium of refractive
index $n$, and separated by a gap, of thickness $d$, filled with a medium of refractive index $n_1<n$ (see Fig. 1).
We refer  the space to a system, $S$, of Cartesian coordinates
$x,y,z$ (unit vector {\bf i, j, k}),
with the $xy$ plane coinciding with the input boundary of the layer, and the
{\bf k} axis normal to the boundaries.
Let us consider a monochromatic plane wave, impinging onto the layer, which
 can be written as\cite{tor}

\begin{equation}
u^i= A^i \exp [ik_0n (\alpha x+\beta y+\gamma z)] \exp (-i\omega t)\:,
\label{wave}
\end{equation}
where $A_i$ is the amplitude at the origin $O$, $k_0=\omega /c$ ($\omega$ is
the angular frequency of the wave),
and $\alpha ,\:\beta ,\gamma$ are the directions cosines.
Without loss of generality, from here on, we shall omit the temporal factor
and we assume $A^i=1$.

\begin{figure}[t]
\begin{center}
\includegraphics[width=.5\textwidth]{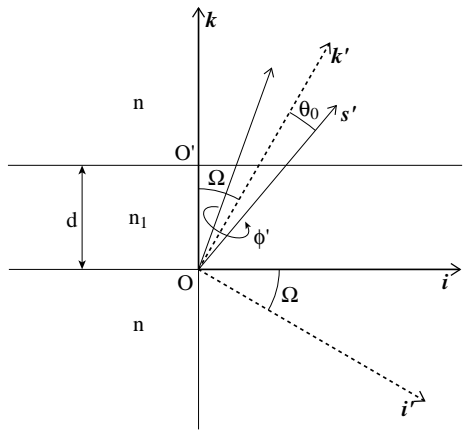}
\end{center}
\caption{Slab, of finite thickness $d$, of a medium with refractive index
$n_1$, surrounded on both sides by a different medium with refractive
index $n>n_1$. The reference system $S$, of unit vectors {\bf i, j, k},
is shown together with the system $S'$, rotated of an angle $\Omega$
with respect to $S$. The vector {\bf s}$^{\prime}$
represents  the direction of
propagation
of a single plane wave. The contribution of all the plane waves, making
the same angle $\theta_0$ with the $z'$ axis, generates a Bessel beam
characterized
by the axicon angle $\theta_0$ and whose axis of symmetry is coincident
with $z'$. The angle $\phi '$ is the azimuthal angle in the plane $x'y'$.}
\label{fig1}
\end{figure}

In Refs. \cite{atti} and \cite{mug00} it has been shown that,
within the same reference system, the  transmitted $u^t$ and the reflected $u^r$ plane waves can be expressed as

\begin{eqnarray}
u^t &=& T \exp [ik_0n(\alpha x+\beta y+\gamma (z-d)]\\
\label{ut}
u^r &=& R \exp [ik_0n(\alpha x+\beta y-\gamma z)] \:,
\label{ur}
\end{eqnarray}
where $T$ and $R$ are the given by

\begin{eqnarray}
T &=& \frac{4inn_1 \gamma\Gamma}
{e_2(n\gamma +in_1 \Gamma )^2-e_1(n\gamma - in_1\Gamma )^2}
\label{t} \\
R &=& \frac{T} {2n_1\Gamma}
\left[ n_1 \Gamma \left( e_1+e_2\right)
+in\gamma  \left( e_1-e_2\right) \right]
-1\:,
\label{r}
\end{eqnarray}
with

\begin{eqnarray}
e_1&=&\exp (-k_0n_1\Gamma d) \:,\nonumber \\
e_2&=& \exp (k_0n_1\Gamma d) \:, \\
\Gamma^2&=&\frac{1}{n_1^2}
\left( n^2-n_1^2-n^2\gamma^2 \right)\:. \nonumber
\end{eqnarray}
The quantity $T$ represents the complex amplitude
of the transmitted wave  at $O'$, and $R$ that of the reflected
wave at $O$.
Note that both $T$ and $R$ depend only on $\gamma$
and not on $\alpha$ and $\beta$.

The above expressions hold when the angle of incidence $i$
onto the layer is larger than the critical angle $i_0$:
this implies that $\sin i > n_1/n$, or, since $\cos i=\gamma$,
that $\gamma^2 <1-(n_1^2/n^2)$ and $\Gamma$ is real.

Let us now consider a set of plane waves, of
the type given in Eq. (\ref{wave}),
having the same complex amplitude
at $O$ and whose directions of propagation form the same
angle $\theta_0$ with the axis {\bf k}.

As usual, we can write the direction cosines as
\begin{eqnarray}
\alpha &=& \sin\theta_0 \cos\phi  \nonumber \\
\beta &=& \sin\theta_0 \sin\phi  \\
\gamma &=& \cos\theta_0 \:,  \nonumber
\end{eqnarray}
where $\phi$ is the azimuthal angle in $xy$ plane.
The superposition of all these waves creates a Bessel beam,
$U_B$, whose axis, {\bf k}$_b$, is normal to the boundaries of the layer
({\bf k}$_b \equiv$ {\bf k})\cite{mug01},
\begin{equation}
U_B^i= J_0(k_0n\rho\sin\theta_0)\exp (ik_0n\cos\theta_0 z)\:,
\label{bes}
\end{equation}
where $\rho$ represents the cylindrical coordinate
in the plane normal to the axis of the beam ($\rho^2 =
x^2+y^2$).
The Bessel beam is, therefore, expressed as an expansion in plane waves,
the directions of propagation of which cover a conical surface of half angle
$\theta_0$ (axicon angle).
Thus, since each plane wave impinges onto the layer with the
same incidence angle $\theta_0$, the refraction angle
$\theta^\prime \: (\sin\theta^\prime=n\sin\theta_0 /n_1)$
will also be the same for all the waves.
Consequently, the transmitted plane waves turn out to have the same
complex amplitude at $O^\prime$, and the transmitted beam
is found to be\cite{mug01,sha}
\begin{equation}
U_B^t= T J_0(k_0n\rho \sin\theta_0)\exp [ik_0n(z-d)\cos\theta_0]\:.
\label{best}
\end{equation}
Analogously, the reflected plane waves turn out to have the
same complex amplitude at $O$, and their
integration over $\phi$ constitutes the reflected Bessel
beam:
\begin{equation}
U_B^r= R J_0(k_0n\rho \sin\theta_0)\exp (-ik_0n z \cos\theta_0)\:.
\end{equation}

Let us consider now the case of oblique incidence, that is
when {\bf k}$_b =${\bf k}$' \neq$ {\bf k}\cite{note}.
In this case, it is expedient
to introduce another reference system, $S^\prime$, whose axis
{\bf k}$'$ is rotated in the $xz$-plane of a given angle
$\Omega$ with respect to {\bf k}\ (see Fig. 1).
The Cartesian axes
{\bf i$^\prime$, j$^\prime$, k$^\prime$} of $S^\prime$
are related to {\bf i, j, k} by

\begin{eqnarray}
{\bf i}^\prime &=& \gamma_k{\bf i}-\alpha_k {\bf k} \nonumber \\
{\bf j}^\prime &=& {\bf j}
\label{i1} \\
{\bf k}^\prime &=& \alpha_k{\bf i}+\gamma_k {\bf k} \:, \nonumber
\end{eqnarray}
where $\alpha_k=\sin\Omega ,\:\gamma_k=\cos\Omega$.

The incident Bessel beam now consists of a set of plane waves, all forming the same angle $\theta_0$ with {\bf k}$'$.
This implies that the plane-wave component impinges onto the first interface of the slab with
different angles of incidence.
It turns out that the incidence
angle varies from $\theta_0 -\Omega$ to $\theta_0 +\Omega$
when $\phi$ ranges from 0 to 2$\pi$.

The generic incidence vector {\bf s$^\prime$} can be written as
\begin{equation}
{\bf s'} = \alpha^\prime {\bf i'}+\beta^\prime {\bf j'}+\gamma^\prime
{\bf k'}
\label{s1}
\end{equation}
with
\begin{eqnarray}
\alpha^\prime &=& \sin\theta_0 \cos\phi^\prime  \nonumber \\
\beta^\prime &=& \sin\theta_0 \sin\phi^\prime
\label{alfa1} \\
\gamma^\prime &=& \cos\theta_0 \:,  \nonumber
\end{eqnarray}
where $\phi^\prime$ is the azimuthal angle in plane $x^\prime
y^\prime$.
Thus, the incidence field due to a generic wave at a given point
$P$ of the space is given by
\begin{equation}
u^i_{ob} = A^i \exp [ik_0n {\bf s'} \cdot \vec{OP}] \:,
\label{wave1}
\end{equation}
and, since $\vec{OP} = x{\bf i} +y{\bf j}  +z{\bf k} $ (or $
= x'{\bf i'} +y'{\bf j'}  +z'{\bf k'} $), by considering
Eqs. (\ref{i1})-(\ref{alfa1}),   we obtain

\begin{equation}
u^i_{ob} =  \exp \{ ik_0n [\alpha_{\phi^\prime }x
+ \beta_{\phi^\prime } y
+ \gamma_{\phi^\prime } z] \} \:,
\label{wave2}
\end{equation}
where

\begin{eqnarray}
&&\alpha_{\phi^\prime }= \gamma_k\sin\theta_0\cos\phi'+
\alpha_k\cos\theta_0  \nonumber \\
&&\beta_{\phi^\prime }= \sin\theta_0\sin\phi' \nonumber \\
&&\gamma_{\phi^\prime }=\gamma_k\cos\theta_0 -
\alpha_k\sin\theta_0\cos\phi' .
\end{eqnarray}
For the reflected $u^r_{ob}$ and transmitted $u^t_{ob}$
fields, we have
\begin{eqnarray}
u^r_{ob} &=& R_{ob} \exp \left\{ ik_0n [\alpha_{\phi '} x +
\beta_{\phi '} y -\gamma_{\phi '} z] \right\}
\label{ur1} \\
u^t_{ob} &=& T_{ob} \exp \left\{ ik_0n [\alpha_{\phi '}x + \beta_{\phi '} y+
\gamma_{\phi '}(z-d)] \right\} \:.
\label{ut1}
\end{eqnarray}
The reflection $R_{ob}$ and transmission $T_{ob}$
coefficients can be still expressed by means of Eqs. (\ref{t}) and
(\ref{r}), with $\gamma$ replaced by $\gamma_{\phi^\prime }$  everywhere.
Consequently, $T_{ob}$ and $R_{ob}$ depend on $\phi'$ as well:

\begin{eqnarray}
T_{ob} &=& \frac{4inn_1 \gamma_{\phi^{\prime} }\Gamma_{\phi^{\prime}}}
{e_{2\phi '} [n\gamma_{\phi^{\prime} } +in_1 \Gamma_{\phi^{\prime} } ]^2
-e_{1\phi '} [n\gamma_{\phi^{\prime} }- in_1\Gamma_{\phi^{\prime} } ]^2}
= |T_{\phi'}| \exp (i\phi_T) \label{t1} \\
R_{ob} &=& \frac{T_{\phi '}}{2n_1\Gamma_{\phi '}}
\left[\frac{}{} n_1
\Gamma_{\phi '}\left( e_{1\phi '}+e_{2\phi '}\right)
+in\gamma_{\phi ' }  \left( e_{1\phi '}-e_{2\phi '}\right) \right]
-1 = |R_{\phi'}| \exp (i\phi_R) \:,
\label{r1}
\end{eqnarray}
where
\begin{equation}
\Gamma_{\phi^{\prime} } = \frac{1}{n_1}\sqrt {n^2(1- {\gamma_{\phi^{\prime}}}^2 )
-{n_1}^2},\:\:\:\:\: e_{1\phi '} = \exp (-k_0 n_1\Gamma_{\phi '}d),\:\:e_{2\phi '}=1/e_{1\phi '}.
\end{equation}

\begin{figure}[t]
\begin{center}
\includegraphics[width=.7\textwidth]{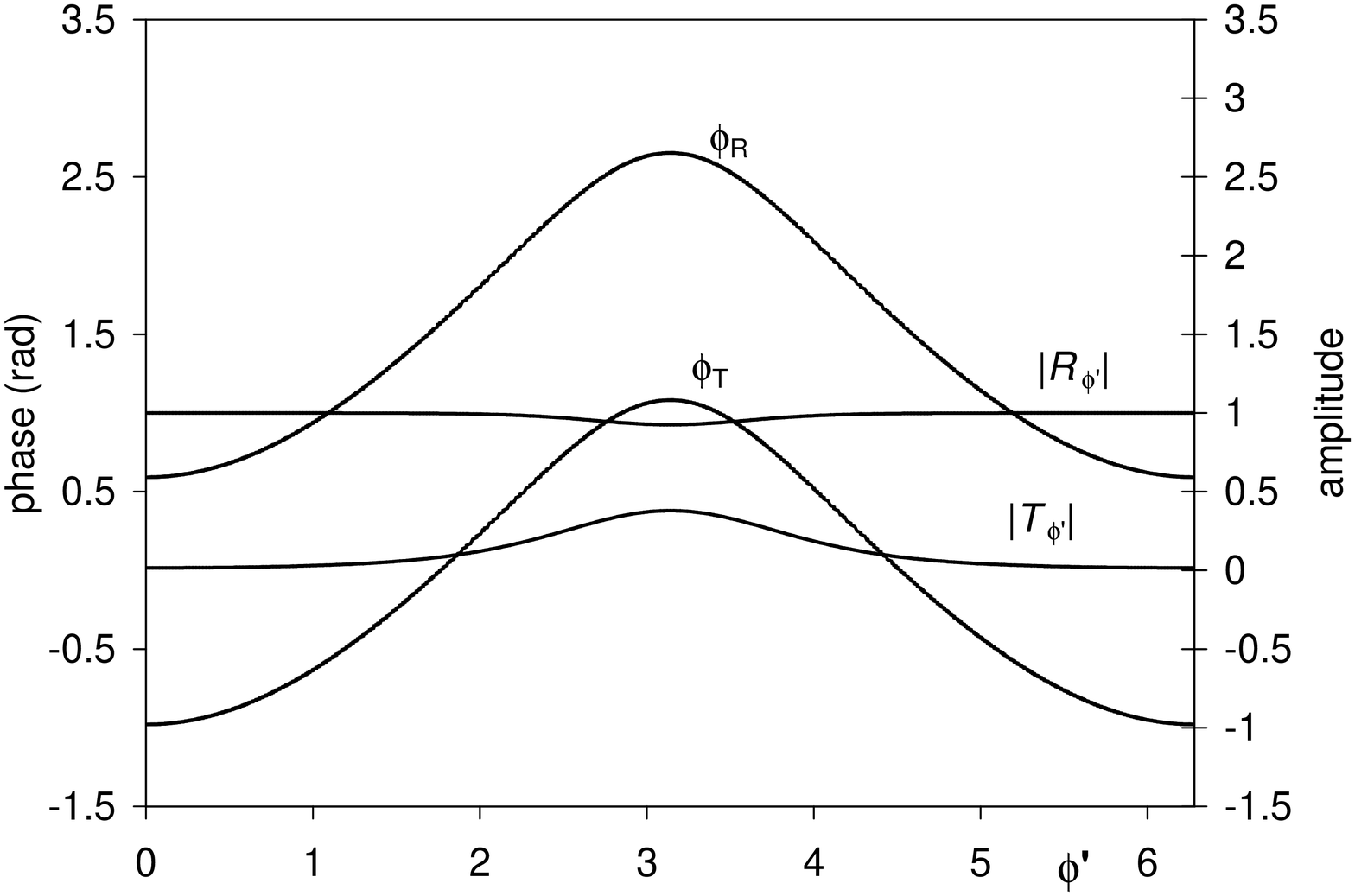}
\end{center}
\caption{Phases and amplitudes of the reflection, $R_{ob}$,
and transmission, $T_{ob}$,
coefficients, as a function of the azimuthal angle $\phi'$.
Parameter values are: $\omega = 60$ rad/s, $n=1.5,\:n_1=1,\:
\Omega\simeq 18^\circ \:
(\alpha_k=0.3),\:\theta_0=60^\circ ,\: d=2$ cm. For this value of $\theta_0,
 \: \Omega =18^\circ$ represents
the limit angle of rotation: for higher values
of $\Omega$, not all the waves forming the beam are in total reflection.}
\label{fig2}
\end{figure}

The amplitudes $|T_{\phi'}|$ and $|R_{\phi'}|$ and phases
$\phi_T$ and $\phi_R$, of $T_{ob}$ and $R_{ob}$, respectively,
are shown in Fig. 2 as a function of $\phi'$, for
$\theta_0 =60^\circ ,\:n_1=1,\:n=1.5$ and $\Omega \simeq 18^\circ \:
(\alpha_k=0.3)$.
With this choice of parameter values, all the waves forming the
beam are in total reflection since the incidence angle
of each wave is larger than the critical angle $i_0=
\sin^{-1}(1/n)= 41.8^\circ$.
For $\theta_0 =60^\circ$, the value of $\Omega \simeq 18^\circ$ represents
the maximum angle possible in order to have total
reflection at the first interface.

In order to find the reflected and transmitted beams, we have to integrate
Eqs. (\ref{ur1}) and (\ref{ut1}) over $\phi'$, between 0 and $2\pi$.
In Figs. 3 and 4, we show the results of the numerical
integration for both the reflected and
transmitted fields as a function of the $x$ coordinate,
for $\theta_0=60^\circ$ and
for normal ($\Omega =0$) and
oblique incidences ($\Omega \simeq 6^\circ ,\: 11^\circ ,\: 18^\circ  ,\:
\alpha_k = 0.1,\: 0.2,\: 0.3$, respectively).
We note that, for $\Omega \neq 0$,
the transmitted field is still characterized
by a main maximum and secondary maxima and minima ,
but suffers deformation with respect to the incident field
and tends to lose its localization\cite{note1}.

\begin{figure}[htb]
\begin{center}
\includegraphics[width=.7\textwidth]{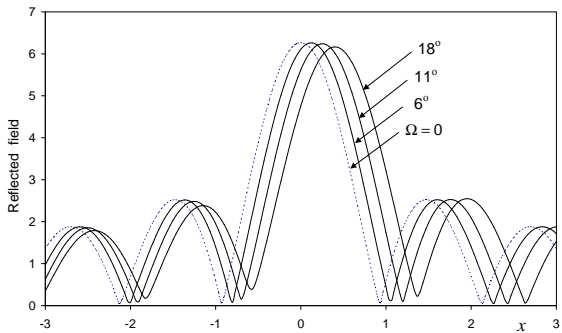}
\end{center}
\caption{Reflected field as a function of the $x$ coordinate, for
normal, $\Omega = 0$ (dashed line), and oblique incidences, $\Omega \simeq 6^\circ ,\:
11^\circ , \:18^\circ \: (\alpha_k=0.1,\:0.2,\:0.3$, respectively).
The reflected field is derived,
for $y=0$ and $z=0$,
by numerical integration of Eq. (17) over $\phi'$, between 0 and $2\pi$.
Other parameter values are as in Fig. 2. The displacement of the field profile, with respect to the normal incidence, evidences the Goos-H\"{a}nchen effect.}
\label{fig3}
\end{figure}

Looking at Figs. 3 and 4, we note that, for the same value of $\Omega$, the reflected field suffers less deformation with respect to the transmitted one.

The deformation of the emerging  Bessel beam (Fig. 4), with respect to the incident one, can be followed by analyzing the field inside the slab
(optical tunneling region).
To this end, let us start again with a single plane wave.
We recall that, in the absence of the second half-space
($d=\infty$), the propagation
after the first
surface is due to evanescent waves which propagate
parallel to the slab.
The presence of the second boundary at $z=d$  originates  anti-evanescent
(or regressive) waves, and the superposition of the two waves, as is well-known,
makes the Poynting vector different from zero also in
the direction perpendicular to the slab.

\begin{figure}[htbp]
\begin{center}
\includegraphics[width=.7\textwidth]{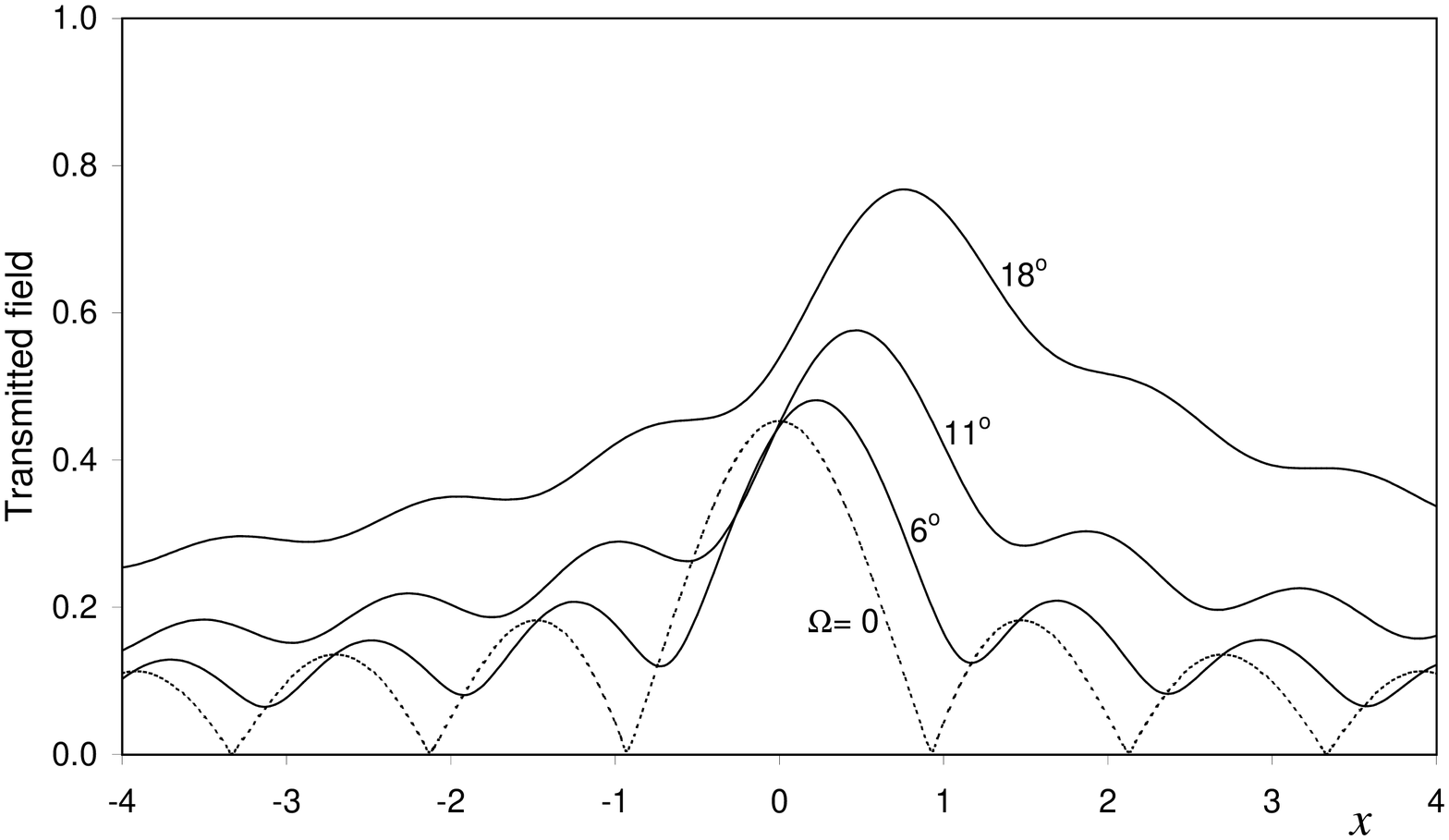}
\end{center}
\caption{Transmitted field, as a function of the $x$ coordinate,
obtained by numerical integration of Eq. (18) over $\phi'$, at $y=0$
and $z=d+1$.
Other parameter values are as in Fig. 3.}
\label{fig4}
\end{figure}

The progressive $u^+$ and regressive $u^-$ waves within the slab
can be written as
\begin{eqnarray}
u^+_{ob} &=& p_{\phi^{\prime} }
\exp \{ ik_0 [n(\alpha_{\phi^{\prime} }x +
\beta_{\phi^{\prime} }y)
+in_1\Gamma_{\phi^{\prime} }z] \}
\label{upiu}  \\
u^-_{ob} &=& r_{\phi^{\prime} } \exp \{ ik_0 [n (\alpha_{\phi^{\prime} }x +
\beta_{\phi^{\prime} } y)
-in_1\Gamma_{\phi^{\prime} }z] \} \:.
\label{umeno}
\end{eqnarray}

\noindent
with
\begin{eqnarray}
p_{\phi '} &=&  \frac{e_{2\phi '}}{2n_1\Gamma_{\phi '}}
(n_1\Gamma_{\phi '} -in\gamma_{\phi '}) T_{\phi '}
\nonumber \\
r_{\phi '} &=& \frac{e_{1\phi '}}{2n_1\Gamma_{\phi '}}
(n_1\Gamma_{\phi '} +in\gamma_{\phi '} ) T_{\phi '} \:.
\label{prfi}
\end{eqnarray}
Equations (\ref{prfi}) were obtained
from the continuity conditions for the tangential component
of both the electric and magnetic fields across the two boundaries,
at $z=0$ and $z=d$\cite{atti}.

\begin{figure}
\begin{center}
\includegraphics[width=.7\textwidth]{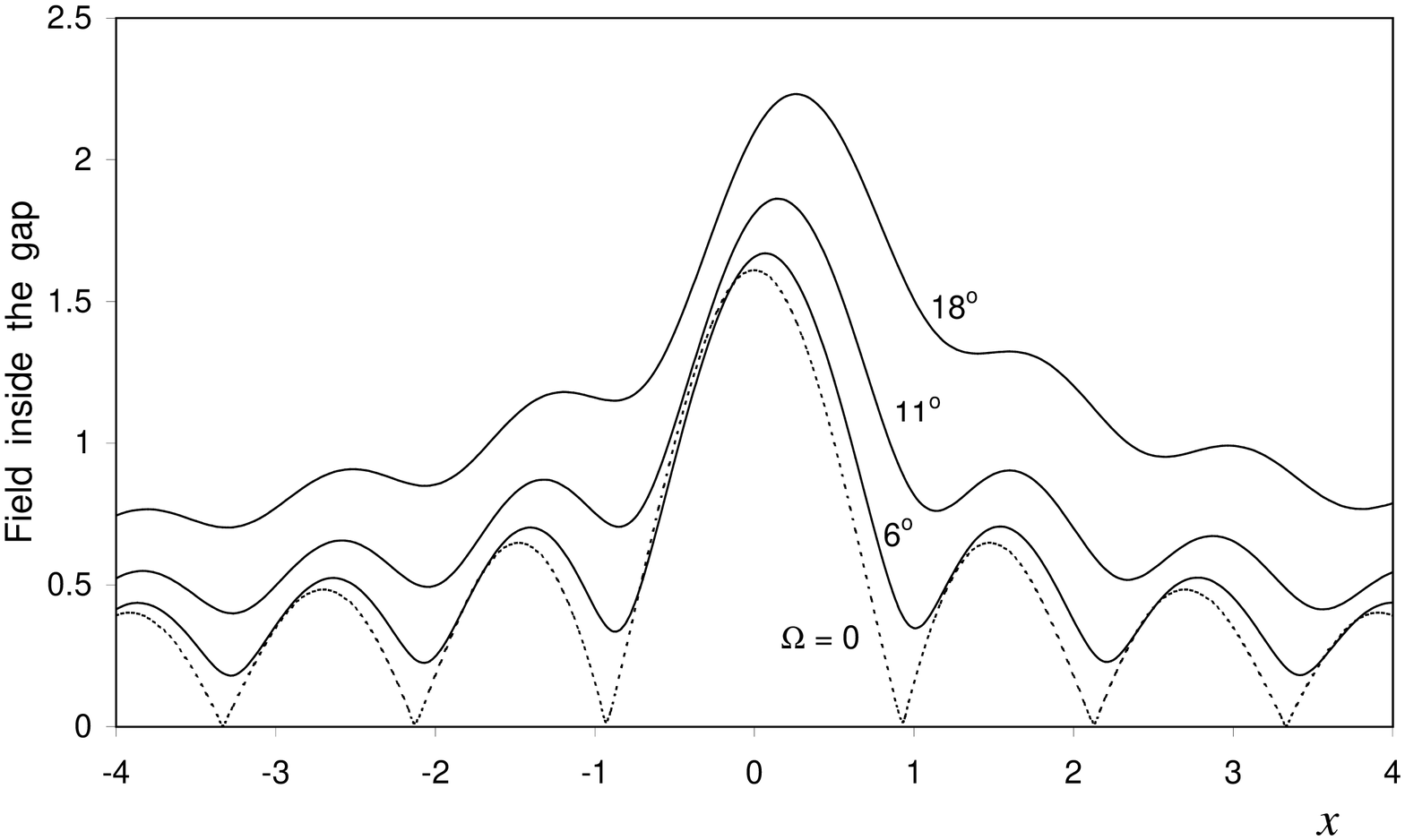}
\end{center}
\caption{Total field inside the gap. The field, given by the
superposition of progressive and regressive waves, was obtained by
numerical integration of Eq. (26) over $\phi'$, at $y=0$ and
$z=d/2$. Other parameter values are as in Fig. 3.} \label{fig5}
\end{figure}

\begin{figure}
\begin{center}
\includegraphics[width=.7\textwidth]{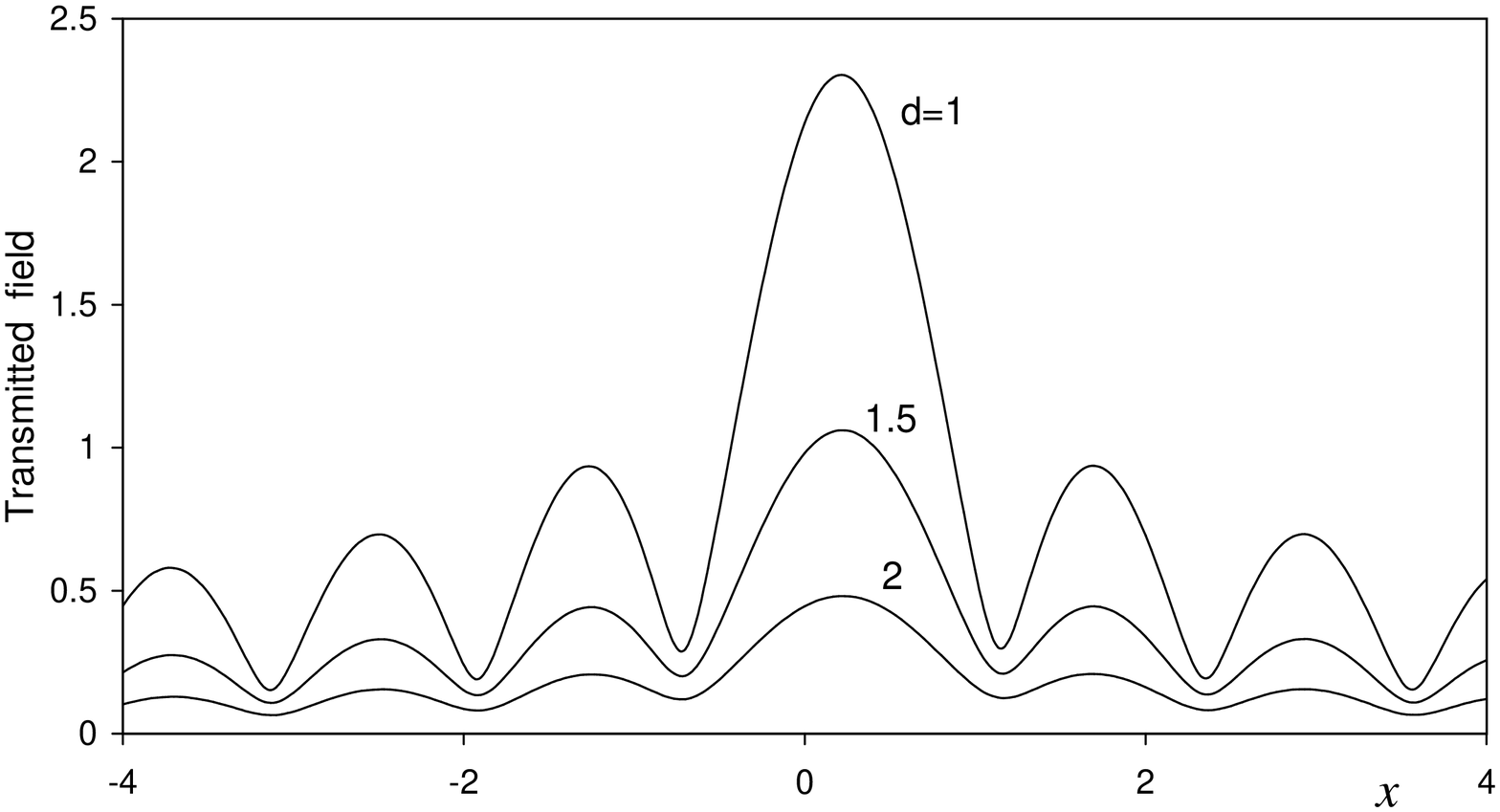}
\end{center}
\caption{Transmitted field obtained by numerical integration of
Eq. (18),
 as a function of the $x$ coordinate, for $\Omega =6^\circ$ and for other parameter values as in Fig. 4.
With these values of the parameter, the amplitude of the field
suffers a decreasing of a factor 5, by varying $d$ from 1 to 2,
while it does not suffer appreciable modification in its shape.}
\label{fig6}
\end{figure}

The field $u_{tot}^g$ inside the gap is given by the superposition of progressive $u^+$ and regressive $u^-$ waves of Eqs.
(\ref{upiu}) and (\ref{umeno}), and can be expressed ad

\begin{equation} 
u_{tot}^g=
\exp [ i( k_0n(\alpha_{\phi^{{\prime}}}x+\beta_{\phi^{\prime}}y)]
\,|T_{\phi^{\prime}}| e^{i\phi_T}
\frac{1}{n_1\Gamma_{\phi'}}
\left[ |A^g| e^{i\eta(z)} \right]
\label{utot}
\end{equation}
where $|T_{\phi^{\prime}}|$ and $\phi_T$ can be derived from
Eq. (\ref{t1}).
The quantities $|A^g|$ and $\eta (z)$, in Eq. (\ref{utot}),
are given by

\begin{eqnarray}
& &|A^g|= \left\{ n_1^2\Gamma_{\phi '}^{2}+(n^2-n_1^2)
\sinh^2[k_0n_1\Gamma_{\phi '} (z-d)]\right \}^{1/2} \nonumber  \\
& &\eta (z) = \arctan \left\{ \frac{n\gamma_{\phi '}}{n_1\Gamma_{\phi '}}\,
\tanh [k_0n_1\Gamma_{\phi '} (z-d)] \right\} \:.
\end{eqnarray}
Again, in order to derive the total Bessel beam we have to integrate the total
field (\ref{utot}) over $\phi '$, from 0 to $2\pi$.
The amplitude of the Bessel beam inside the gap, at $z=d/2$ and $y=0$,
is shown in Fig. 5 as a function of the $x$ coordinate, for three different values of the incidence angle $\Omega$.
We note that, for oblique incidence,
the Bessel beam  starts to lose its characteristic property of
localized wave in the passage through the slab.
In Fig. 6, we show the behavior of the transmitted field as a function of the
$x$ coordinate, for three different values of the slab's thickness.
We note that the dependence on $d$ produces a
strong variation in the amplitude value, with no appreciable effect in the shape of the field: the maxima and minima positions remain unchanged.
The same behavior holds also for the reflected and internal fields, with the only difference lying in the fact that the variation in the amplitude is inappreciable for the reflected field, while the amplitude is halved in the field inside the slab.

The numerical analysis was performed for an axicon angle of 60$^\circ$
in order to have a clearer evidence of the
delocalization effect due to the  passage through the slab. For smaller
axicon angles, the incidence angle must also be smaller in order to have total reflection, and the
effect of deformation due to oblique incidence is very poor.

An interesting aspect, related to the subject treated here,
is the analysis of the wavefronts of the beam inside
the slab, in order to have informations about the
direction of propagation and the
phase velocity in the tunneling region.
This kind of analysis, however, is beyond the scope of the present work and will be reported elsewhere.

\vspace{1 cm}
\noindent
{\bf Acknowledgements}

Special thanks are due to Laura Ronchi Abbozzo for
useful discussions and suggestions.


\begin{thebibliography}{50}
\bibitem{note0}
The use of the word ``beam'' is accepted because the
$U_B$ field is well
limited to a restricted zone of space
(the first zero of $J_0$) along the radial
coordinate.
\bibitem{note1}
However, it may be verified that the transmitted amplitude ($z>d$)
propagates rigidly in the {\bf k}$'$ direction (the axis of the beam)
with no modification in
its shape. An analogous property also holds for the reflected field.
\end{thebibliography}
\end{document}